\documentclass[reprint,showpacs,superscriptaddress]{revtex4-1}
\usepackage{amsmath}
\usepackage{amsfonts}
\usepackage{amssymb}
\usepackage{graphicx}
\usepackage{braket}
\usepackage{hyperref}

\newcommand{\doo}{\delta\omega}
\newcommand{\doc}{\delta\omega_\mathrm{c}}

\newcommand{\Dc}{\Delta_\mathrm{c}}

\newcommand{\Dee}{\Delta_\mathrm{e}}

\renewcommand{\gg}{g_\mathrm{\gamma}}
\newcommand{\go}{\gamma_\mathrm{0}}
\newcommand{\gc}{g_\mathrm{c}}
\newcommand{\gct}{\tilde{g}_\mathrm{c}}
\newcommand{\gk}{g_\kappa}
\newcommand{\gkt}{\tilde{g}_\kappa}
\newcommand{\ggt}{\tilde{g}_\gamma}

\newcommand{\kk}{\kappa}
\newcommand{\ko}{\kappa_\mathrm{0}}

\newcommand{\ksub}{\kappa_\mathrm{sub}}
\newcommand{\kint}{\kappa_\mathrm{int}}
\newcommand{\kref}{\kappa_\mathrm{ref}}

\newcommand{\oo}{\omega}
\renewcommand{\oc}{\omega_\mathrm{c}}

\newcommand{\um}{\mathrm{\mu m}}
\newcommand{\im}{\mathrm{i}}

\newcommand{\tens}{\stackrel{\leftrightarrow}}

\newcommand{\al}{\tens{\alpha}}
\newcommand{\E}{\mathbf{E}}
\newcommand{\dE}{\delta\mathbf{E}}

\newcommand{\Eo}{\E_0}

\renewcommand{\H}{\mathbf{H}}
\newcommand{\dH}{\delta\mathbf{H}}
\newcommand{\Ho}{\H_0}
\renewcommand{\r}{\mathbf{r}}

\newcommand{\ee}{\epsilon}
\newcommand{\eo}{\epsilon_0}
\newcommand{\mo}{\mu_0}
\newcommand{\De}{\Delta \epsilon}
\newcommand{\Dm}{\Delta \mu}

\newcommand{\p}{\mathbf{p}}

\newcommand{\DV}{\Delta V}
\begin{document}
\title{Perturbing open cavities: Anomalous resonance frequency shifts in a hybrid cavity-nanoantenna system}
\author{Freek Ruesink}
\affiliation{FOM Institute AMOLF, Science Park 104, 1098 XG, Amsterdam, The Netherlands}
\author{Hugo M. \surname{Doeleman}}
\affiliation{FOM Institute AMOLF, Science Park 104, 1098 XG, Amsterdam, The Netherlands}
\affiliation{Institute of Physics, University of Amsterdam, P.O. Box 94485, 1090 GL Amsterdam, The Netherlands}
\author{Ruud Hendrikx}
\affiliation{FOM Institute AMOLF, Science Park 104, 1098 XG, Amsterdam, The Netherlands}
\author{A. Femius \surname{Koenderink}}
\affiliation{FOM Institute AMOLF, Science Park 104, 1098 XG, Amsterdam, The Netherlands}
\affiliation{Institute of Physics, University of Amsterdam, P.O. Box 94485, 1090 GL Amsterdam, The Netherlands}
\author{Ewold Verhagen}
\email{verhagen@amolf.nl}
\affiliation{FOM Institute AMOLF, Science Park 104, 1098 XG, Amsterdam, The Netherlands}
\pacs{42.82.Fv, 78.67.-n, 42.25.Hz, 42.25.Fx}
%\date{\today}
\begin{abstract}
The influence of a small perturbation on a cavity mode plays an important role in fields like optical sensing, cavity quantum electrodynamics and cavity optomechanics.
Typically, the resulting cavity frequency shift directly relates to the polarizability of the perturbation.
Here we demonstrate that particles perturbing a radiating cavity can induce strong frequency shifts that are opposite to, and even exceed, the effects based on the particles' polarizability.
A full electrodynamic theory reveals that these anomalous results rely on a non-trivial phase relation between cavity and nanoparticle radiation, allowing back-action via the radiation continuum.
In addition, an intuitive model based on coupled mode theory is presented that relates the phenomenon to retardation.
Because of the ubiquity of dissipation, we expect these findings to benefit the understanding and engineering of a wide class of systems.
\end{abstract}

\maketitle
The fact that a small perturbation of a potential can influence the distribution of a system’s energy levels is a well-known principle permeating various branches of physics.
In quantum mechanics, for example, the effect of a perturbing potential $H'$ on an eigenstate $\ket{\psi_0}$ is that it modifies its unperturbed energy $U_0$ by an amount $\delta U=\braket{\psi_{0}|H'|\psi}$, where $\ket{\psi}$ is the new eigenstate and we assume $\braket{\psi_0|\psi}\approx1$. In electrodynamics, a local change of potential (\emph{i.e.} permittivity) can impact the frequency of a resonant cavity.
This is at the basis of many applications that use the influence of a perturbing atom, molecule, or dielectric body to establish an interaction that can be exploited for optical sensing or control~\cite{Raimond2001,Lodahl2015,Vollmer2008,Fan2008,Aspelmeyer2014,Zhu2010,Thompson2013,Goban2014}.
The shift of a mode's complex eigenfrequency $\oo=\oc-\im\kappa/2$, with cavity resonance frequency $\oc$ and linewidth $\kappa$, due to a local permittivity perturbation $\Delta\epsilon$ contained in a volume $\Delta V$ is given by $\doo/\oo = - \int_{\DV} dV \left[\eo\De\E_0^\ast\cdot\E_p\right]/4U_0$.
Here $\E_0$ and $U_0$ represent the field and total energy of the unperturbed cavity mode, respectively, and $\E_p$ is the perturbed field~\cite{Waldron1960, Bethe1943}.
In particular, if $\Delta V$ is small enough such that the perturbing particle can be described in the dipole approximation, the complex frequency shift is directly related to the particle \emph{polarizability} $\alpha$, reading $\doo/\oo = - \alpha |\E_0|^{2}/4U_{0}$~\cite{Waldron1960}.
However, this widely employed result of Bethe-Schwinger perturbation theory~\cite{Bethe1943} is strictly only valid when radiation to the far field is negligible.
In several recent developments, radiation loss proved decisive in determining a system's eigenmode, \emph{e.g.} for so-called `states bound in the continuum'~\cite{Friedrich1985,Hsu2013,Monticone2014} and in describing complex plasmonic resonators~\cite{Taubert2012,Liu2009}.
In that context, the question arises to what extent the conventional paradigm to determine perturbed cavity frequencies holds in practical, open, systems.

\begin{figure}[t]
\center
\includegraphics[width=\linewidth]{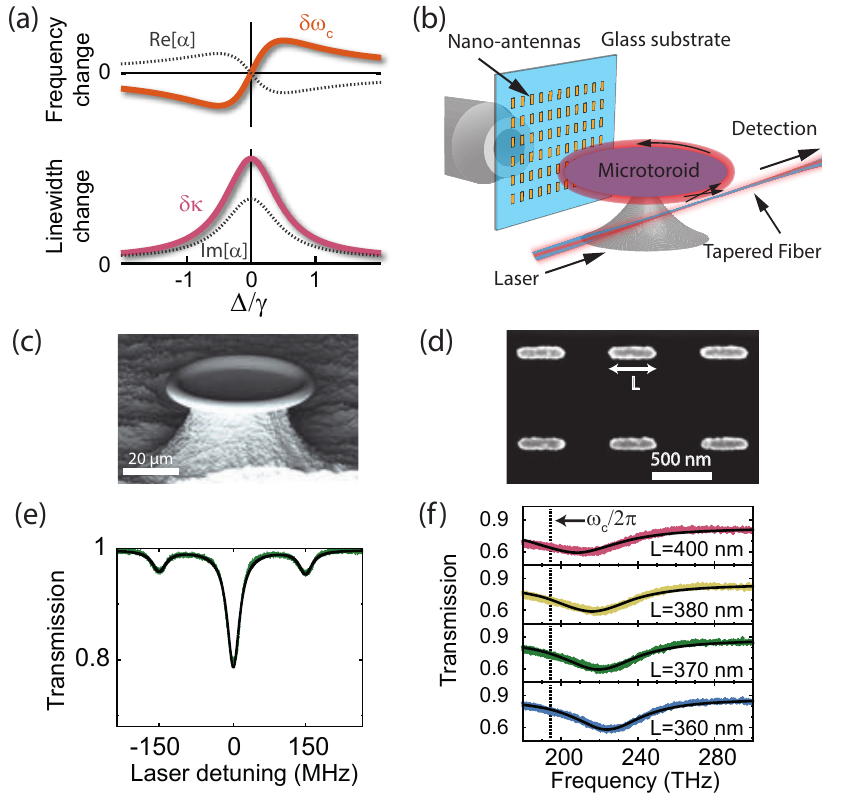}
 \caption{ a) Cavity frequency and linewidth change due to perturbation by a small resonator, considering only the resonator polarizability. Dashed lines indicate the real (top) and imaginary (bottom) part of the polarizability as a function of $\Delta/\gamma$.
b) An array of perturbing gold nanoantennas is placed in the near field of a toroidal microcavity. Light transmission through the tapered fiber is detected to determine the cavity eigenfrequency shift.
c) Scanning electron microscopy (SEM) image of a high-Q silica microtoroid at the edge of a chip.
d) SEM image of gold nanoantennas on a glass substrate.
e) Transmittance through the tapered fiber around a cavity mode resonance. A Lorentzian fit (black solid line) is used to determine the optical linewidth ($\approx$30 MHz) and frequency of the cavity. The sidebands result from a 150-MHz phase modulation used to calibrate the frequency axis.
f) Normal-incidence transmission spectra of gold nanoantenna (design length $L$) arrays. Lorentzian fits (sold lines) give the antenna resonance frequency and linewidth. The dashed line indicates the cavity frequency.
}
 \label{fig:BS_system}
 \end{figure}

Here, we study the eigenfrequencies of a radiating optical cavity as it is perturbed by carefully designed resonant plasmonic nanoparticles.
For a resonant perturbation with center frequency $\omega_{\mathrm{a}}$ and linewidth $\gamma$, the value of $\alpha$ strongly depends on the detuning $\Delta=\oc-\omega_{\mathrm{a}}$.
In absence of radiation, the cavity mode thus redshifts(blueshifts) at negative(positive) values of $\Delta$ and the linewidth broadens near $\Delta=0$, in direct response to Re[$\alpha$] and Im[$\alpha$], respectively (Fig~\ref{fig:BS_system}a).
Tuning $\omega_{\mathrm{a}}$ by varying the length of the (plasmonic) resonators allows to systematically study the induced cavity response as a function of detuning.
Importantly, our experiment is designed such that radiation losses from the cavity mode and plasmonic resonators overlap. This, as we will show, leads to a strong additional contribution to $\doo$ which is not captured by $\alpha$.
We study this `radiation interaction', as we will call the effect, in detail and show that it can induce strong eigenfrequency shifts that are opposite to, and even exceed, the effects based on the particles' polarizability.
%Figure 1

A sketch of the system is shown in Fig.~\ref{fig:BS_system}b.
The experiments are performed using a fundamental cavity mode (194.4 THz, Q$\sim$6.5$\times10^{6}$, TE polarized) of a toroidal silica microcavity fabricated on the edge of a silicon chip ($\approx$36 $\mu$m diameter, Fig.~\ref{fig:BS_system}c)~\cite{Armani2003, Anetsberger2009}.
The cavity is perturbed by gold nanoantennas deposited on a glass substrate (Fig.~\ref{fig:BS_system}d), which are controllably placed in the evanescent field of the cavity.
The antennas (nanorods of length $L$, width 120 nm and thickness 40 nm) are aligned with their (long) principal dipole axis to the polarization of the cavity mode.
A frequency-swept narrowband laser source ($\sim$0.7 $\mu$W) is coupled into a tapered fiber that is brought close to the cavity.
The transmission spectrum through the fiber shows a Lorentzian dip around the cavity resonance frequency (Fig.~\ref{fig:BS_system}e), from which we determine the (perturbed) resonance frequency $\oc$ and linewidth $\kappa$ of the cavity mode.
Independent normal-incidence transmission measurements (Fig.~\ref{fig:BS_system}f) on the fabricated antenna arrays, in absence of the cavity, yield the normalized cavity-antenna detuning $\Delta/\gamma$.
Notably, the presence of the glass substrate allows the cavity mode to radiate into the glass at a well defined angle just beyond the critical angle and, similarly, allows scattering of antenna radiation into the substrate~\cite{Novotny2012}.
As the antennas are coherently excited by the cavity field, from which also the cavity radiation originates, the zeroth-order diffraction by the array is expected to overlap with the cavity radiation.
%Figure 2
\begin{figure}[tb]
\center
\includegraphics[width=\linewidth]{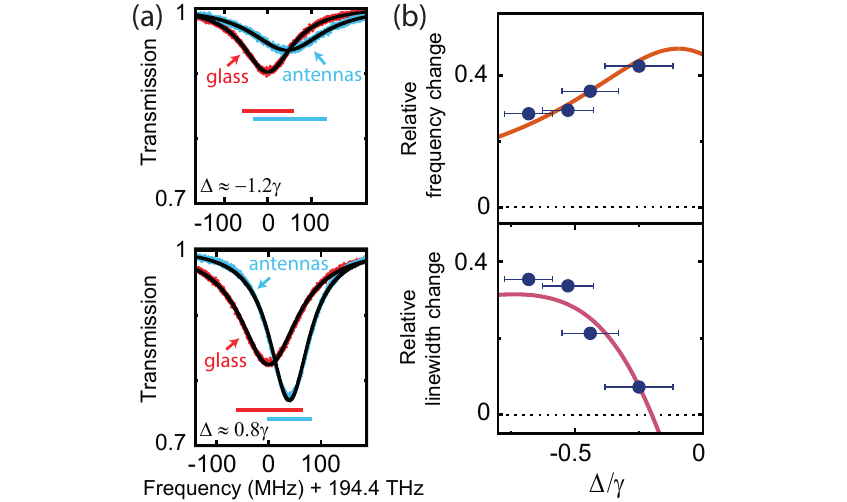}
    \caption{a) Depending on $\Delta$, antenna arrays induce linewidth broadening (top) and narrowing (bottom), while simultaneously inducing the cavity resonance frequency to blueshift.
The blue and red horizontal bars represent the linewidths of fitted Lorentzian lineshapes (black lines).
b) Shift in resonance frequency and linewidth of the cavity mode, normalized to its linewidth in absence of the antennas~\cite{SM}, due to perturbation by antenna arrays with a constant pitch size and varying antenna length.
Top: For all detunings a blueshift of cavity resonance frequency is observed.
Bottom: Approaching $\Delta=0$ induces linewidth narrowing. Both the blueshift and linewidth narrowing are contrary to the expectation based on the particles' polarizability (Fig.~\ref{fig:BS_system}a).
Error bars depict standard deviation, vertical error bars fall within the plot markers.
The solid lines represent a coupled mode theory fit.}
    \label{fig:BS_exp}
\end{figure}

To measure the change of the complex cavity resonance frequency due to the nanoantennas, we compare the frequency and linewidth of the cavity mode with and without the antenna array placed in the near-field of the cavity.
Our measurement procedure~\cite{SM} ensures that in both cases the sample is positioned at the same distance to the cavity.
Figure~\ref{fig:BS_exp}a shows the results for two different arrays of nanoantennas: In the top figure, an antenna array ($L=360$, pitch 800(900) nm along the long(short) axes of the antennas) is seen to induce a broadening of the cavity linewidth, together with a \emph{blueshift} of the frequency.
The latter is clearly surprising, given the fact that $\Delta\approx-1.2\gamma$ for this sample~\cite{SM}, where one expects a \emph{redshift} if one only considers the resonant particle's polarizability (Fig.~\ref{fig:BS_system}a).
The bottom panel in Fig.~\ref{fig:BS_exp}a shows that for another array ($L=400$, pitch 800(1100) nm) with slightly positive detuning ($\Delta\approx 0.8\gamma$) again a blueshift is observed, but this time accompanied by \emph{narrowing} of the cavity mode.
Such a reduction of damping can obviously not be ascribed to the particle's polarizability alone, as its imaginary part is necessarily positive (Fig.~\ref{fig:BS_system}a).

To systematically study these effects, we gradually tune the resonance frequency $\omega_{a}$ by varying the length of the antennas (Fig.~\ref{fig:BS_system}f).
The array pitch is kept fixed at 800(1500) nm along the long(short) antenna axes, chosen such that inter-antenna coupling is negligible.
Moreover, in all examples we show, Bragg-scattering between clockwise and counter-clockwise modes~\cite{Mazzei2007} is small enough such that mode splitting induced by the nanoantennas is smaller than the cavity linewidth.
Figure~\ref{fig:BS_exp}b shows the resulting cavity frequency and linewidth changes, normalized to the cavity linewidth on glass to allow averaging multiple scans~\cite{SM}, as a function of $\Delta/\gamma$.
For these negative detunings, a consistent increase of the resonance frequency (i.e. blueshift) is observed that slightly rises as the antenna frequency approaches that of the cavity.
As such, both the sign and the trend of $\doc$ are incongruous with the expectation based on the polarizability (Fig.~\ref{fig:BS_system}a).
Moreover, the broadening of the cavity, observed for a detuning of $\Delta\approx-0.6\gamma$, quickly reduces as the --~strongly scattering~-- antennas are tuned closer to the cavity resonance.
We note that these trends are consistently observed also at other periodicities~\cite{SM} and that the shifts cannot be explained by thermal heating, negligible at the employed powers.

To explore the origin of these surprising results, it is necessary to consider the complete Bethe-Schwinger cavity perturbation equation~\cite{Bethe1943}, derived without neglecting radiation:
\begin{align}
\doo &= - \oo \frac{\alpha |\mathbf{E_{0}}|^{2}}{4 \, U_{0}}  \nonumber\\
&-\frac{i}{4 \, U_{0}} \int_{\delta V} dA \left[(\dE \times \mathbf{H_{0}^{\ast}}) \cdot \mathbf{\hat{n}} +  (\mathbf{E_{0}^{\ast}} \times \dH) \cdot \mathbf{\hat{n}}  \right].
\label{eq:BS}
\end{align}
This full expression, which is also valid in any open, non-Hermitian system~\cite{SM}, contains an additional integral term involving radiated fields at a surface $\delta V$ (with normal unit vector $\mathbf{\hat{n}}$), enclosing the same volume that was used to evaluate the total energy $U_0$.
Here, $\dE$ and $\dH$ are the difference between the fields in the presence ($\E_p$, $\H_p$) and absence ($\E_0$, $\H_0$) of the perturbation, and can as such be associated with the field scattered by the perturbation.
Thus, the cavity eigenfrequency is additionally modified by an energy flux that is evaluated by \emph{combining} the fields of the perturbed and unperturbed eigenmodes; \emph{i.e}, an overlap of scattering by the perturbation and direct cavity radiation.
Remarkably, the integral is independent of the distance at which $\delta V$ is chosen, consistent with the fact that it can be associated with radiation of the system.
To date, this far-field contribution has been omitted in practically all analyses of cavity perturbation. In a select number of experiments, a reduction of cavity linewidth was observed~\cite{Koenderink2005,Robinson2008,Burresi2010} and tentatively attributed to interference of radiation of the cavity and scattering by the perturbation.
Importantly, the integral suggests that in principle also the resonance frequency (i.e., the real part of $\doo$) can be affected by the same mechanism, if the phase difference between cavity and scattered radiation in the far field (captured in $\E_0$ and $\dE$, respectively) differs from 0 or $\pi$.
So far, the frequency has been expected to rely only on the local variation of the applied potential \cite{Vignolini2010,Lalouat2007,Burresi2010, Arnold2003, Vollmer2008, Novotny2012} as it is contained in the first term of (\ref{eq:BS}).
%Figure 3
\begin{figure}[tb]
\center
\includegraphics[width=\linewidth]{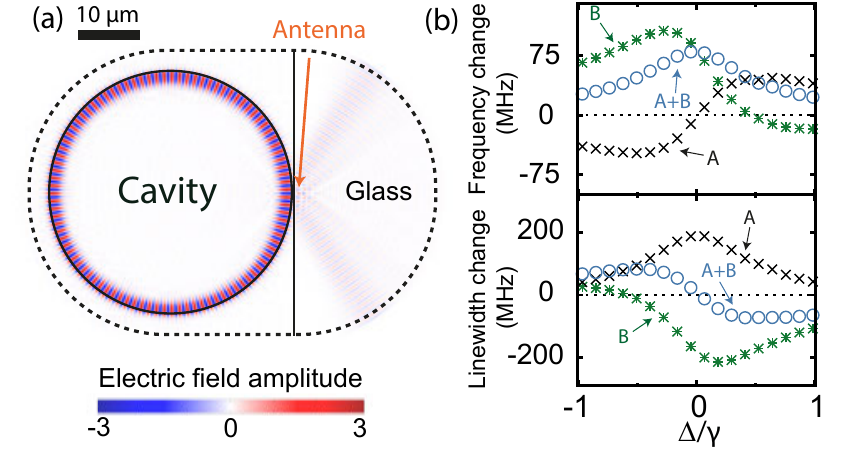}
 \caption{a) Electric field profile in a simulated system containing cavity, antenna and substrate. Both antenna and cavity radiate into the glass under approximately the same angle, resulting in a radiation interaction. The contribution of this interaction is calculated on the dashed black line (integral of (\ref{eq:BS})). The arrow points to the position of the antenna.
b) Top(bottom): the change in resonance frequency(linewidth) calculated using (\ref{eq:BS}). Black crosses (`A') show the contribution of the polarizability to the eigenfrequency shift $\doo$, which resembles the lineshape as we expect it from a resonant particle (Fig.~\ref{fig:BS_system}a). The radiative contribution (green stars, `B') shows distinct blueshifts and linewidth narrowing near $\Delta/\gamma=0$. Adding both terms (blue circles, `A+B') yields the complete cavity eigenfrequency shift, which matches the experimental trends.}
 \label{fig:BS_comsol}
 \end{figure}

To verify the importance of radiation in a complete description of the induced cavity response, we perform finite-element method eigenfrequency studies with and without antenna \cite{SM} in a similar two-dimensional system (Fig.~\ref{fig:BS_comsol}a), from which we extract the fields $\E_0$, $\H_0$, $\dE$ and $\dH$.
This allows the calculation of the individual terms of (\ref{eq:BS}), displayed as a function of detuning in Fig.~\ref{fig:BS_comsol}b.
The first term (black crosses, `A'), contributes to $\doo$ according to the expected behaviour for a perturbing resonator sketched in Fig.~\ref{fig:BS_system}a: it produces a dispersive detuning dependence of $\doc$ and a dissipative trend for $\delta\kappa$.
This immediately disqualifies this term as an explanation for the measured changes in cavity resonance and linewidth.
The second term, \emph{i.e.} the contribution to the change in eigenfrequency related to the radiation interaction (green stars, `B'), shows a dramatically different behavior: it causes distinct blueshifts and linewidth narrowing around $\Delta/\gamma=0$.
Surprisingly, the magnitude of the blueshifts induced by this term can even exceed the contribution due to $\alpha$.
This is a key point of our observation, stressing the importance of this new contribution in a model system.
The sum of both terms (blue circles, A+B) yields the complete eigenfrequency shift of the perturbed cavity, which now qualitatively matches our experimental results; blueshifts and linewidth narrowing around $\Delta/\gamma=0$.
These calculations thus confirm that radiation interactions in open systems can lead to both cavity blueshifts and linewidth narrowing and importantly, that this effect can even \emph{dominate the total change in eigenfrequency} of a cavity mode.
%Figure 4
\begin{figure}[tb]
\center
\includegraphics[width=\linewidth]{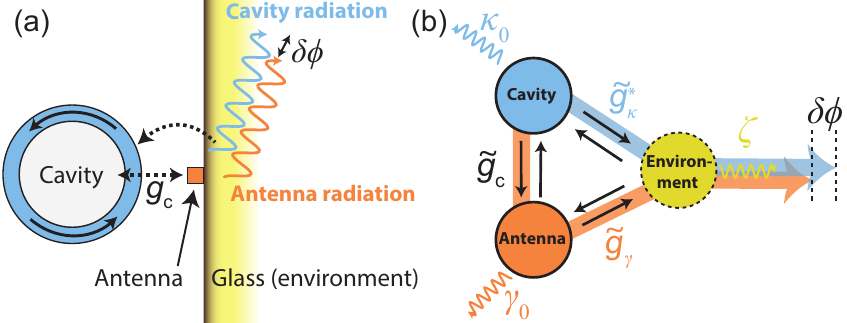}
 \caption{a) The cavity and antenna are coupled with rate $g_\mathrm{c}$, which is the near-field interaction scaling with the polarizability of the antenna, and can radiate into the continuum with a certain delay ($\delta\phi$) with respect to each other. The interference between both radiation profiles and the overlap with the cavity mode results in a back-action effect (single-headed dashed arrow) on the cavity mode.
 b) Coupled-mode theory. The cavity and antenna are coupled with rate $\gct$ and couple to the continuum with complex rates $\gkt = \gk \, e^{\im \phi_{\kappa}}$ and $\ggt = \gg \, e^{\im \phi_{\gamma}}$ respectively. The difference between $\phi_{\kk}$ and $\phi_{\gamma}$, summed with the phase response of the antenna, is now equal to the phase difference $\delta\phi$. $\ko$ and $\go$ are the losses of cavity and antenna into non-coupled channels. The environment is modeled as an oscillator with decay rate $\zeta$.}
\label{fig:BS_coupled_oscillators}
\end{figure}

It is essential to realize that a modification of the cavity resonance frequency (in contrast to the alteration of the linewidth) due to back-action via the radiation continuum only occurs when the phase $\delta\phi$ between cavity and antenna radiation is different from 0 or $\pi$ (Fig.~\ref{fig:BS_coupled_oscillators}a).
The exact strength of this back-action, and as such the sum of direct (\emph{i.e.} due to $\alpha$) and indirect (\emph{i.e.} via the radiation continuum) coupling between cavity and antenna, will depend on the overlap between their radiation profiles and on the value of $\delta\phi$.

This phase difference and its physical connection to back-action can be captured in a simple model (Fig.~\ref{fig:BS_coupled_oscillators}b) based on temporal coupled-mode theory~\cite{Suh2004}.
Let us assume for now that the cavity and antenna are coupled with complex coupling rate $\gct=\gc\, e^{\im \phi_{\mathrm{c}}}$ and can both interact with the environment (glass) with coupling rates $\gkt = \gk \, e^{\im \phi_{\kappa}}$ and $\ggt = \gg \, e^{\im \phi_{\gamma}}$, respectively.
We treat the joint environment as a separate mode to allow backaction, taking the limit of large decay rate $\zeta$ to mimic a broadband continuum.
The cavity and antenna, considered separately, each decay via this continuum at a rate $\kk_{1}=4|\gk|^{2}/\zeta$ and $\gamma_{1}=4|\gg|^{2}/\zeta$, respectively.
This is distinguished from decay into all other, non-overlapping, modes at rates $\ko$ and $\go$, such that $\ko+\kk_{1}=\kappa$ and $\go+\gamma_{1}=\gamma$.
Importantly, the phase difference between cavity radiation (blue path, Fig.~\ref{fig:BS_coupled_oscillators}b) and antenna scattering (orange path, Fig.~\ref{fig:BS_coupled_oscillators}b) in this model now directly relates to $\delta\phi$ via $\delta\phi=\pi+\Phi-\text{arg}[\Delta+\im\gamma/2]$, where $\Phi=\phi_{\mathrm{c}}+\phi_{\kappa}+\phi_{\gamma}$, and $\text{arg}[\Delta+\im\gamma/2]$ is the phase response of the antennas~\cite{SM}.
The natural mode of interest, with complex frequency $\omega$, of this hybrid system can be found by equating the determinant of the coupling matrix~\cite{SM}
\begin{equation}
M=\begin{pmatrix}
\omega-\oc+\im\ko/2 & \gct^\ast & \gkt \\
\gct & \Delta+\im\go/2 & \ggt^\ast \\
\gkt^\ast & \ggt & \im\zeta/2
\end{pmatrix}
\label{eq:BS_matrix}
\end{equation}
to zero.
Solving for complex $\omega$ yields
\begin{equation}
\delta\oc-\im\delta\kappa/2 = \frac{\gc^2-\kk_{1}\gamma_{1}/4 + \im\gc\sqrt{\kk_{1}\gamma_{1}}\cos{\Phi}}{\Delta+\im\gamma/2},
\label{eq:3CO_Domega}
\end{equation}
which depends on the radiation overlap $\kk_{1}\gamma_{1}$, (constant) phase $\Phi$ and the coupling rate $\gc$.
Because the overlap and phase $\Phi$ cannot be independently chosen, it is impossible to directly fit (\ref{eq:3CO_Domega}) to the experimental data.
Therefore, we retrieve $\delta\phi$ from the finite element simulation and, using the relation between $\delta\phi$ and $\Phi$, obtain $\Phi=3.76\times10^{-4}\pi$.
Note that this value of $\Phi$ implies that $\delta\phi\approx\pi/2$ at $\Delta=0$, completely opposite to the conventionally assumed case where destructive interference ($\delta\phi=\pi$) only contributes to the linewidth of the cavity mode.
Fixing $\Phi$ at this value and constraining the experimentally found values of $\kappa$, $\Delta$ and $\gamma$ (retrieved from independent spectroscopic measurements, Fig.~\ref{fig:BS_system}), a fit of (\ref{eq:3CO_Domega}) yields excellent correspondence to our data in Fig.~\ref{fig:BS_exp}b (solid lines), resulting in an overlap of $\kk_{1}\gamma_{1}/\kappa\gamma=0.68$ and coupling rate $\gc/2\pi=21.6$ GHz.

Concluding, we have shown how coupling through the radiation continuum leads to cavity blueshifts and linewidth narrowing in a coupled cavity-antenna system.
Surprisingly, the non-local, radiative effect on the cavity is larger than the induced cavity response due to local perturbations, and is not just an artefact of large radiation losses but can even manifest itself in a high-Q cavity such as studied here.
Therefore, similar effects are expected in many other systems where cavity and perturbation can radiate into the same channels, such as photonic crystals~\cite{Koenderink2005,Robinson2008,Burresi2010,Barth2010} and Fabry-P\'erot etalons~\cite{Mitra2010, Chanda2011,Ameling2013}.
In addition, it was shown that there is a direct link between the phase $\delta\phi$ in the Bethe-Schwinger equation and the contributions to back-action in a coupled-mode model.

Interestingly, these results could also shed new light on radiative interactions in strongly radiating systems such as metamaterials and complex plasmonic resonators.
In~\cite{Taubert2012}, it was postulated that radiation can be included in the coupling between two resonators by using a complex coupling rate, while in \cite{Bakhti2015}, a complex coupling rate was argued to originate from a complex-valued extinction cross-section.
In fact, such a complex coupling rate between two resonators is easily derived from the coupled-mode model presented in this work~\cite{SM}.
Furthermore, it will be interesting to see how the concepts and theory established here relate to recently developed methods to relying on the normalization of leaky modes to describe the response of complex photonic systems~\cite{Sauvan2013,Yang2015}.
In a different context, recent advances in optomechanics using simultaneous dispersive and dissipative (\emph{i.e} radiative) coupling \cite{Sawadsky2015,Hryciw2015,Xuereb2011} have been reported to enable on-resonance cooling~\cite{Sawadsky2015} and sensitive readout of nanomechanical motion~\cite{Hryciw2015}.
As such, we expect our work to be directly important to the design of novel optical devices such as sensors, for example, and benefit the understanding of the physics of open systems in optics and beyond.  \\

This work is part of the research programme of the Foundation for Fundamental Research on Matter (FOM), which is part of the Netherlands Organisation for Scientic Research (NWO). EV gratefully acknowledges support from an NWO Vidi grant.
%%%%
%\bibliography{1_CavityPerturbation}

%merlin.mbs apsrev4-1.bst 2010-07-25 4.21a (PWD, AO, DPC) hacked
%Control: key (0)
%Control: author (8) initials jnrlst
%Control: editor formatted (1) identically to author
%Control: production of article title (-1) disabled
%Control: page (0) single
%Control: year (1) truncated
%Control: production of eprint (0) enabled
%

\newpage

%%%%%%%%%%%%%%%%%
%		SI		 %
%%%%%%%%%%%%%%%%%
%SI commands for renumbering
\renewcommand{\theequation}{S\arabic{equation}}
\renewcommand{\thefigure}{S\arabic{figure}}
\onecolumngrid
\begin{center}
\Large
{\textbf{Supplementary Material}}
\end{center}
\normalsize
\section{Bethe-Schwinger cavity perturbation in open systems}
\label{sec:BS}
In this section, we show the derivation of the exact Bethe-Schwinger cavity perturbation formula, analogous to the approach by \cite{SIWaldron1960,SIBethe1943}.
Afterwards, we discuss how small approximations result in equation (\ref{eq:BS}) of the main text.
We note, that the calculation of the different contributions to the eigenfrequency change in our simulations did not involve the approximated equation~(\ref{eq:BS}), but the exact formula as shown in equation~(\ref{eq:Bethe}).

\subsection{Eigenmodes}
We consider the modes of an open cavity described by a spatial distribution of permittivity $\eo\ee(\r)$ and permeability $\mo\mu(\r)$ (in the following, the spatial dependence of both is implicitly assumed). The eigenmodes of the system are found by solving Maxwell's equations in all of space in the absence of external drives. This yields solutions of the form
\begin{eqnarray}
\E(\r,t)=\Eo(\r) e^{-\im\oo t},\\
\H(\r,t)=\Ho(\r) e^{-\im\oo t},
\end{eqnarray}
which satisfy Maxwell's equations for some complex frequency
\begin{equation}
\oo=\oc-\im \frac{\kk}{2},
\end{equation}
where $\oc$ and $\kk$ denote the (real) resonance frequency and energy decay rate, respectively. Typically, we will consider systems where $\kk\ll\oc$.
Important relationships between $\Eo$ and $\Ho$ include
\begin{eqnarray}
\nabla\times\Eo = \im\oo\mo\mu\Ho && \nabla\times\Eo^\ast = -\im\oo\mo\mu\Ho^\ast,\\
\nabla\times\Ho = -\im\oo\eo\ee\Eo && \nabla\times\Ho^\ast = \im\oo\eo\ee\Eo^\ast.
\end{eqnarray}

\subsection{Dielectric perturbation}
We now consider that the cavity is perturbed, such that in a finite volume $\DV$, the relative permittivity and permeability are changed to new values
\begin{eqnarray}
\ee_p=\ee+\De,\\
\mu_p=\mu+\Dm.
\end{eqnarray}
This perturbed system has different eigenmodes written as
\begin{eqnarray}
\E(\r,t) = \E_p(\r) e^{-\im\oo_p t} = \left(\Eo(\r)+\dE(\r)\right) e^{-\im\left(\oo+\doo\right)t},\\
\H(\r,t) = \H_p(\r) e^{-\im\oo_p t} = \left(\Ho(\r)+\dH(\r)\right) e^{-\im\left(\oo+\doo\right)t},
\end{eqnarray}
where $\E_p(\r)$ and $\oo_p$ describe the spatial dependence and resonance frequency of the perturbed cavity mode. Defining $\dE=\E_p-\Eo$, $\dH=\H_p-\Ho$ and $\doo=\oo_p-\oo$ yields
\begin{eqnarray}
\nabla\times\left(\Eo+\dE\right) = \im\left(\oo+\doo\right)\mo\mu_p\left(\Ho+\dH\right),\\
\nabla\times\left(\Ho+\dH\right) = -\im\left(\oo+\doo\right)\eo\ee_p\left(\Eo+\dE\right).
\end{eqnarray}
Combining the above equations for the curl of the fields gives
\begin{eqnarray}
\nabla\times\dE = \im\oo_p \mo\mu_p\H_p - \im\oo\mo\mu\Ho,\\
\nabla\times\dH = -\im\oo_p \eo\ee_p\E_p + \im\oo\eo\ee\Eo.
\end{eqnarray}
We next take the dot product of $\Ho^\ast$ and $\Eo^\ast$ with the curls of $\dE$ and $\dH$, respectively and rewrite both making use of the vector identity
\begin{equation}
\mathbf{a}\cdot\left(\nabla\times\mathbf{b}\right) = \mathbf{b}\cdot\left(\nabla\times\mathbf{a}\right) - \nabla\cdot\left(\mathbf{a}\times\mathbf{b}\right).
\end{equation}
Subtracting the obtained expressions gives us
\begin{align}
\doo\left(\eo\ee\Eo^\ast\cdot\E_p + \mo\mu\Ho^\ast\cdot\H_p\right) =
-\left(\oo+\doo\right)&\left(\eo\De\Eo^\ast\cdot\E_p+\mo\Dm\Ho^\ast\cdot\H_p\right)  \nonumber\\
&-\im\left(\nabla\cdot\left(\dE\times\Ho^\ast\right)+\nabla\cdot\left(\Eo^\ast\times\dH\right)\right).
\end{align}
Finally, we take the integral of both sides over a (very large) volume $V$, and apply Gauss's theorem to arrive at the \emph{Bethe-Schwinger equation}~\footnote{Due to the different (engineering) convention used in COMSOL, implementation of this formula requires to switch the sign in front of the surface integral from (-) to (+).}:
\begin{align}
\doo\int_V dV \left[\eo\ee\Eo^\ast\cdot\E_p + \mo\mu\Ho^\ast\cdot\H_p\right] =
-\left(\oo +\doo\right)& \int_{\DV} dV \left[\eo\De\Eo^\ast\cdot\E_p+\mo\Dm\Ho^\ast\cdot\H_p\right] \nonumber  \\
&- \im \int_{\partial V} dA \left[\left(\dE\times\Ho^\ast\right)\cdot\hat{\mathbf{n}}+\left(\Eo^\ast\times\dH\right)\cdot\hat{\mathbf{n}}\right],
\label{eq:Bethe}
\end{align}
where $\partial V$ denotes the boundary of $V$ and $\hat{\mathbf{n}}$ the unit vector normal to that surface.

\subsection{Approximating the cavity perturbation}
The above equation (\ref{eq:Bethe}) yields an \emph{exact} expression for the (complex) cavity shift
\begin{equation}
\doo = \doc-\im\frac{\delta\kk}{2}.
\end{equation}
To arrive at equation (\ref{eq:BS}) of the main paper, we start by making the following approximations: If the effect of the perturbation is small, then $\oo+\doo\approx\oo$, and when the total volume is large compared to the volume of the perturbation, it will be allowed to omit the contribution of $\Eo^\ast\cdot\dE$ in the integral on the left side of equation (\ref{eq:Bethe}), which can now be written as
\begin{align*}
\doo\int_V dV \left[\eo\ee\Eo^\ast\cdot\E_p + \mo\mu\Ho^\ast\cdot\H_p\right] \approx
\doo\int_V dV \left[\eo\ee\left|\Eo\right|^2 + \mo\mu\left|\Ho\right|^2\right]=4\doo U_0,
\end{align*}
where $U_0$ denotes the energy stored in the unperturbed cavity. Now the Bethe-Schwinger equation becomes
\begin{equation}
\doo = -\frac{\oo}{4U_0} \int_{\DV} dV \left[\eo\De\Eo^\ast\cdot\E_p+\mo\Dm\Ho^\ast\cdot\H_p\right]\\
 -\frac{\im}{4U_0} \int_{\partial V} dA \left[\left(\dE\times\Ho^\ast\right)\cdot\hat{\mathbf{n}}+\left(\Eo^\ast\times\dH\right)\cdot\hat{\mathbf{n}}\right].
\label{bethenew}
\end{equation}
If the perturbation is caused by a variation of the permittivity ($\Delta\mu=0$) and if the dimensions of its volume $\DV$ are small compared to the scale over which the cavity field $\Eo$ varies, one can make a useful approximation by considering the perturbation as a polarizable (point) dipole. Its total dipole moment $\p$ is given by
\begin{equation}
\p=\int_{\DV} \mathrm{d}V \frac{\mathrm{d}\p}{\mathrm{d}V},
\end{equation}
where we introduced the dipole moment per unit volume, which we identify as
\begin{equation}
\frac{\mathrm{d}\p}{\mathrm{d}V}(\r) = \eo\left(\ee_p-\ee\right)\E_p(\r) = \eo\De\E_p(\r).
\end{equation}
Taking into account that $\Eo$ can be considered constant over $\DV$, we can now write the integral over $\DV$ in equation (\ref{bethenew}) as
\begin{equation}
\Eo^\ast\cdot\eo\int_{\DV}\mathrm{d}V\De\E_p = \Eo^\ast\cdot\p = \Eo^\ast\cdot\al\Eo,\label{intalpha}
\end{equation}
where we have introduced the polarizability $\al$ defined in this case through
\begin{equation}
\p = \al\Eo.
\end{equation}
If $\al$ is diagonal, and we denote the magnitude of the polarizability in the direction of the cavity field by the scalar $\alpha$, the integral over $\DV$ reduces to $\alpha\left|\Eo\right|^2$.
In this approach, we arrive at a new version of the Bethe-Schwinger equation for the perturbation of the cavity frequency:
\begin{equation}
\doo = -\oo\frac{\alpha\left|\Eo\right|^2}{4U_0}
 -\frac{\im}{4U_0} \int_{\partial V} dA \left[\left(\dE\times\Ho^\ast\right)\cdot\hat{\mathbf{n}}+\left(\Eo^\ast\times\dH\right)\cdot\hat{\mathbf{n}}\right],
\label{bethealpha}
\end{equation}

Here we see how the real part of the particle's polarizability contributes to a frequency shift of the cavity, but so does the imaginary part of the surface integral. Likewise, the change of the cavity linewidth is related to the imaginary part of $\alpha$ and the real part of the integral over $\partial V$.

\section{Experimental details}
\label{sec:BS_experimental}
\subsection{Sample fabrication}
Gold nanoantennas are fabricated in an array on a 170 $\mu$m thick glass coverslide.
To start, a 130 nm layer of ZEP-520 resist is spin-coated on top of the coverslide.
The nanoantennas are patterned into the resist using electron-beam lithography.
After development, thermal evaporation of gold and a lift-off step yield the desired antennas.
The antenna width and thickness were designed to be 120 and 40 nm and the length was varied between 400 and 360 nm. The pitch along the short-axes of the antennas was varied between 750 and 1500 nm, and the pitch along the long-axes was fixed at 800 nm.
A high-Q silica microtoroid (diameter $\approx$ 36 $\um$) is fabricated on the edge of a silicon sample, largely following methods as previously reported \cite{SIArmani2003, SIAnetsberger2009}.
In this work, spin-coating (ma-N 2410) and subsequent cleaving of the sample enabled targeted e-beam lithography of the disks. \\

\subsection{Experimental Setup}
\begin{figure}[t]
\center
  \includegraphics[width=0.6\linewidth]{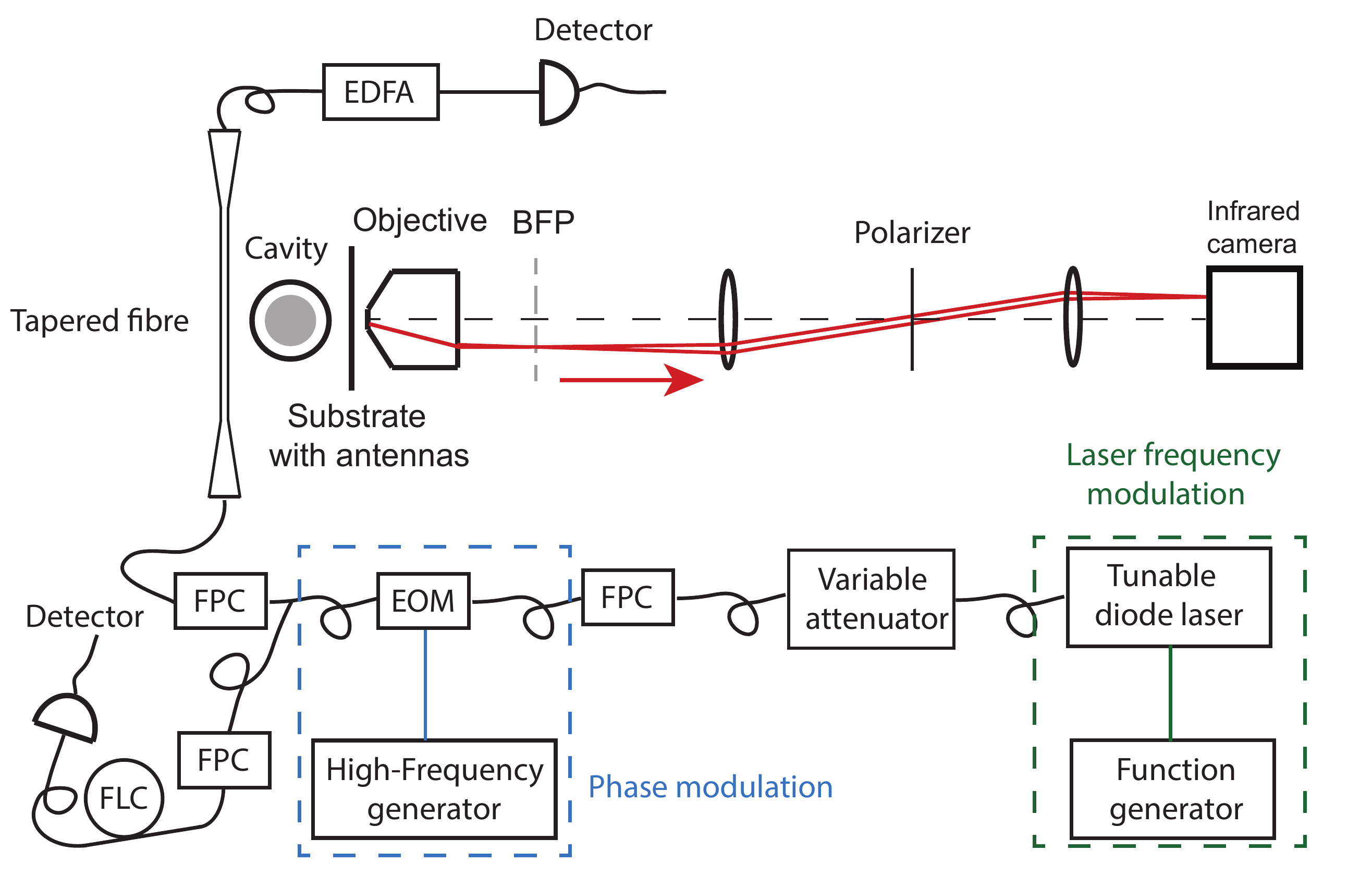}\\
  \caption{\textbf{Experimental setup.} See text for details.}
  \label{fig:setup}
\end{figure}
\begin{figure}[t]
\center
  \includegraphics[width=0.3\linewidth]{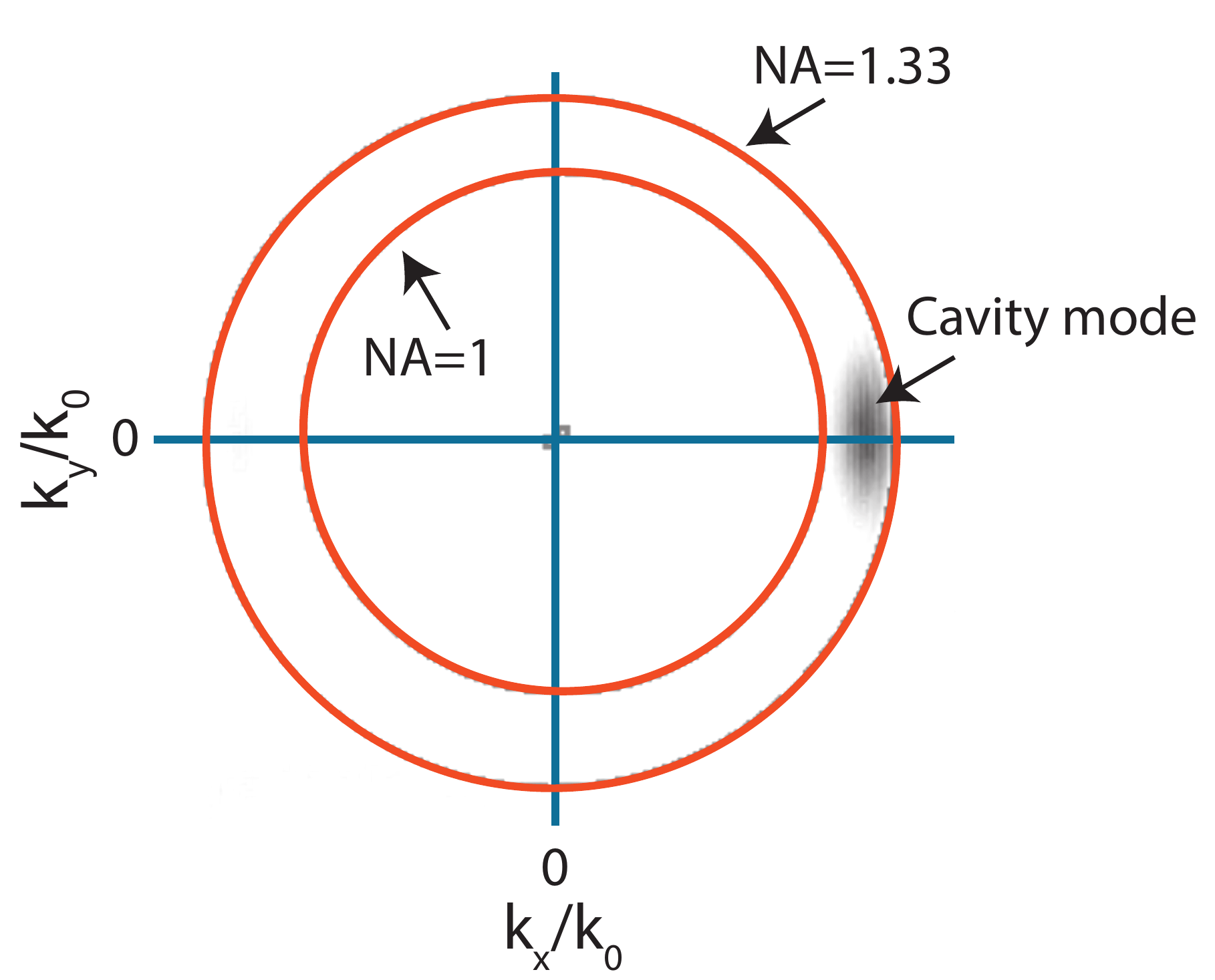}\\
  \caption{\textbf{Mode characterization.} Illumination of the BFP of our objective with white light (not shown) is used to calibrate the image on our infrared camera. This involves fitting the NA = 1 ring and maximum collection angle of our objective (orange circles), which are imaged on the camera. Sample tilt results in a slight offset between the outer and NA=1 ring. Subsequent monitoring of the radiation leakage of our cavity mode enables its characterization (fundamental TE mode) and an estimate (see text) of the effective mode index ($\approx 1.35$). The blue solid lines serve as a guide to the eye.}
  \label{fig:mode}
\end{figure}

Figure~\ref{fig:setup} shows the experimental setup used in this work. We use a tunable fiber-coupled external cavity diode laser (New Focus TLB-6728, $<$100 kHz linewidth) to probe the microcavity.
The laser frequency is controlled by an external voltage from a function generator.
Coupling of light into the cavity is achieved using a tapered optical fiber, of which the position is controlled using piezo positioners (not shown).
The electro-optical modulator (EOM, EOspace) is used to generate sidebands of the cavity mode at known RF frequency, which allows for calibration of our optical frequency axis.
Fiber polarization controllers (FPC) ensure effective coupling into the optical cavity mode of interest.
Before detection, the light that is transmitted through the tapered fiber is amplified using an Erbium Doped Fiber Amplifier (EDFA).
Radiation leakage from the microcavity is collected using an objective (Nikon, CFI Apo TIRF 100x) to obtain information regarding polarization and mode profile (through fourier-space imaging and polarization analysis of the back focal plane (BFP), Fig.~\ref{fig:mode}) of the cavity mode.
Using this method we identify a fundamental cavity mode, of which the polarization is aligned with the long axes of the antennas.
Due to limitations with regards to the numerical aperture (NA) of our objective (NA $\approx$ 1.33), imaging of the complete cavity mode-profile is impossible, thus resulting in a cut-off of the imaged mode profile.
Considering this cut-off at NA=1.33 and the refractive index of silica (1.46), we estimate an effective mode index of around 1.35.
We check the frequency stability of the laser itself using a fiber loop reference cavity (FLC).
\begin{figure}[t]
\center
  \includegraphics[width=0.5\linewidth]{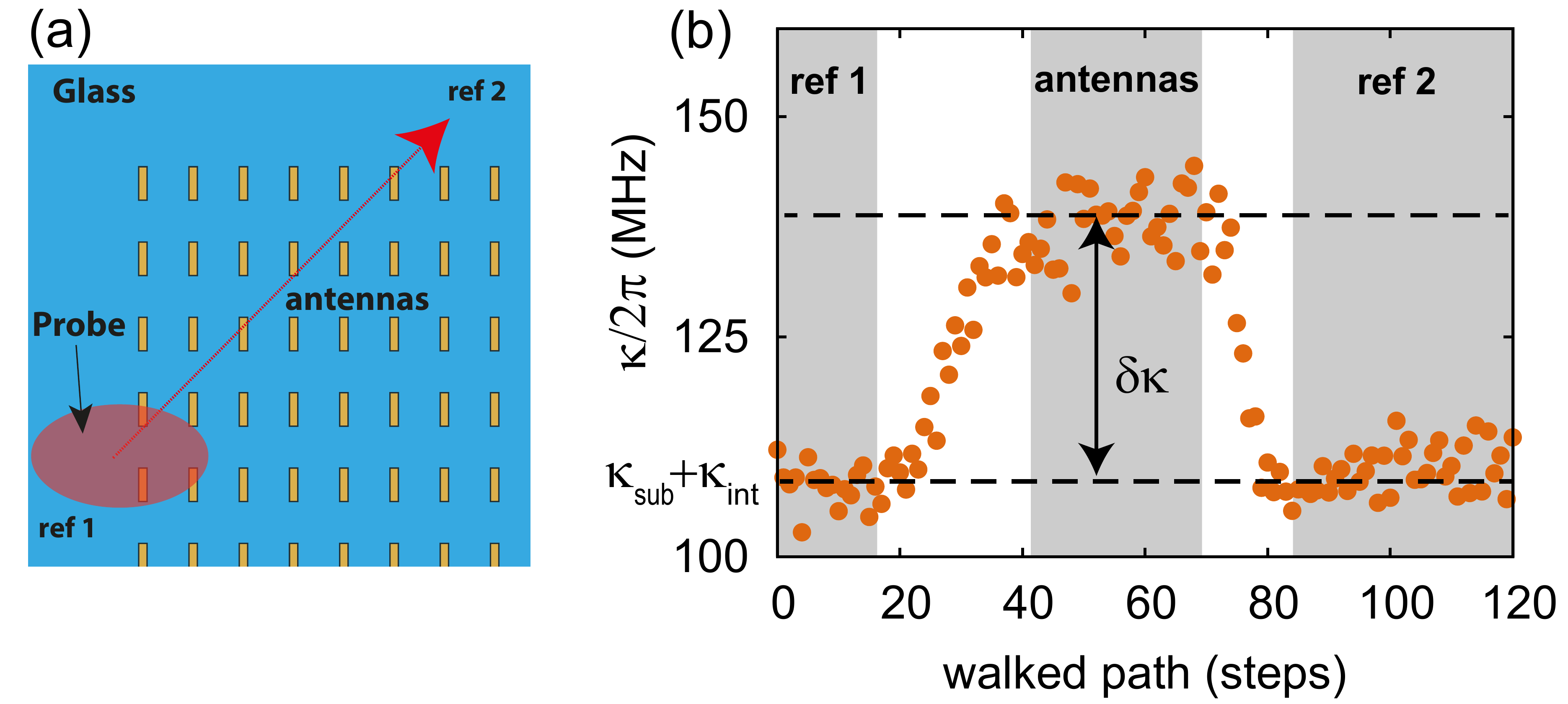}
  \caption{\textbf{Scanning along an array.} a) Cartoon of a scan of the microcavity along the antenna array. Careful alignment allows us to scan over the antennas by moving our microcavity between two predefined points (ref 1 and ref 2), which serve as our reference measurements on glass. Between those points we measure the effect of the antennas on the cavity mode. b) Evolution of the cavity linewidth during a scan along the antenna array, where each point is retrieved from a fit with a Lorentzian lineshape. When scanning along the antennas, we make sure that we see a \emph{plateau} of data points. In our analysis, we only include the data points indicated by the grey regions, thus excluding the data taken during transit between reference and antennas. For the calculation of $\delta\kk$, we average all the points clearly on the \emph{plateau} and subtract the average reference value on glass ($\kref = \ksub+\kint$).}
  \label{fig:sweep}
\end{figure}
\subsection{Measurement procedure}
Figure~\ref{fig:sweep}a shows that the experimental data is obtained by stepwise scanning the microcavity, which is placed on a piezo-electric stage, along an antenna array and a stretch of bare substrate at a distance of $\sim$1.1 $\um$ (based on simulations).
During each scan, the cavity-fiber distance is kept constant, which is checked before and after each experiment when the cavity is positioned at opposite positions adjacent to the antenna arrays.
This check is performed by monitoring the cavity linewidth (broadened by the presence of the substrate), which depends strongly on the distance to the substrate.
Each scan consists of 80-120 steps and multiple scans (2-5) are performed on each array.
The change in resonance frequency ($\doc$) and linewidth ($\delta\kk$) due to the antennas is obtained by subtracting the cavity resonance frequency and linewidth without the array placed in the near-field of the cavity, from the cavity resonance frequency and linewidth with the antenna array placed in the evanescent field of the cavity mode.
Example data for the change in linewidth is shown in Fig.~\ref{fig:sweep}b.
To correct for any remaining sample tilt and/or slow drift, reference measurements on glass taken at the start and end of a scan are averaged.
The values of $\doc$ and $\delta\kk$ are calculated by excluding data points taken during the transition from glass to array.
\begin{figure}[tb]
\center
  \includegraphics[width=0.6\linewidth]{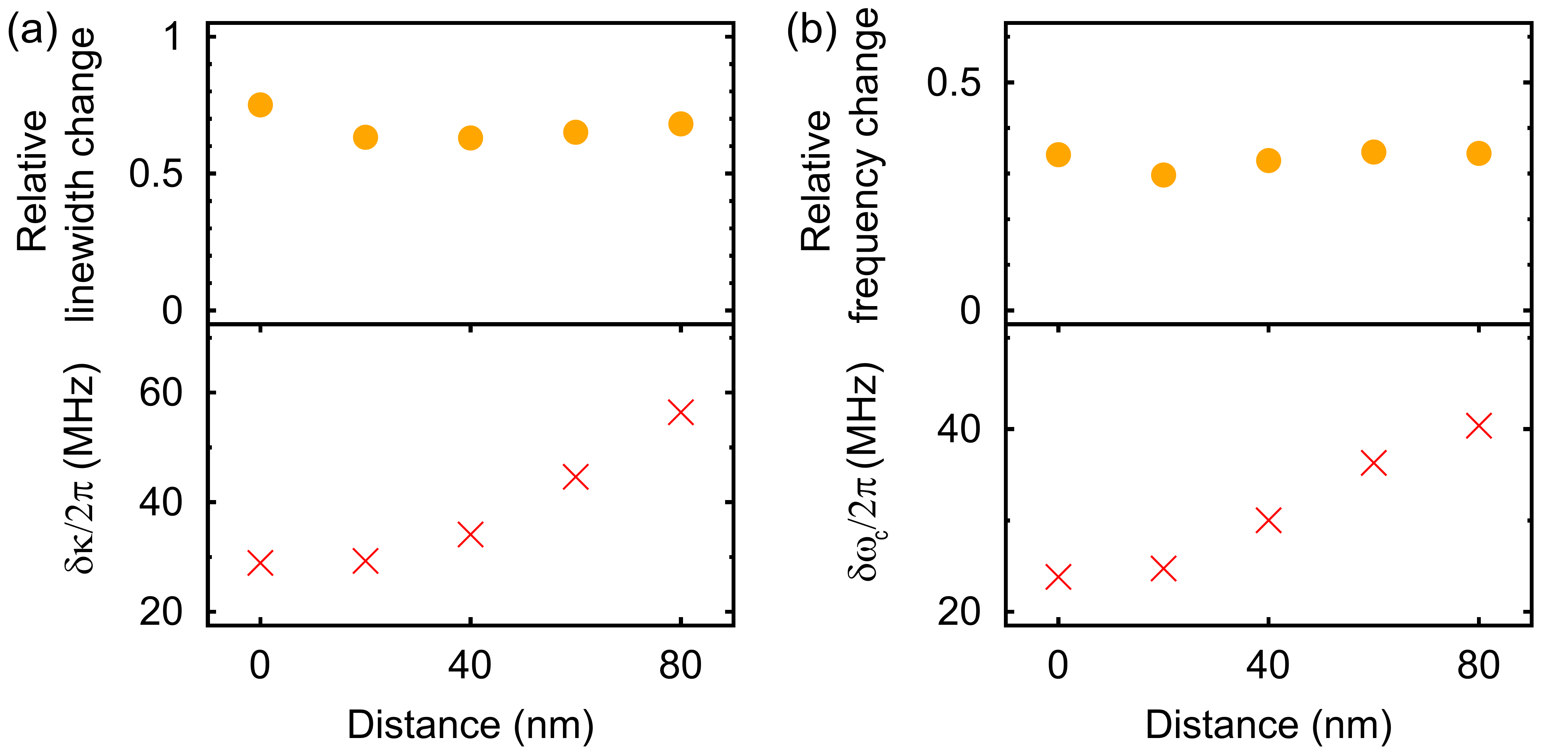}
  \caption{\textbf{Validation of dimensionless parameter.} The measured effect on the cavity linewidth (a, bottom) and resonance frequency (b, bottom) due to the antennas is clearly increasing when the cavity travels a certain distance towards the antennas. As shown in the top panel of (a) and (b), our analysis results in a constant dimensionless quantity for the change in linewidth and frequency which does not depend on the distance to the antenna array.}
  \label{fig:approach}
\end{figure}

Even though we check that all measurements are performed at the same cavity-sample distance, we would like to eliminate any errors due to small variations of that distance. To this end, we retrieve parameters that are robust towards small distance changes by considering the relative change in linewidth(resonance frequency) due to the antennas: $\delta\kk(\doc)/\ksub$, where $\ksub$ is the broadening the cavity mode experiences due to the glass substrate ($\ksub=\kref-\kint$, with $\kint$ the decay rate of the bare cavity).
Figure~\ref{fig:approach} shows that this method results in a relative change of the resonance frequency $\oc$ and linewidth $\kappa$ which is independent of the distance to the substrate. It is obvious that this quantity now allows straightforward averaging over multiple scans on the same array and comparison between different antenna arrays.
Errors on the mean are calculated for $\doc$ and $\delta\kappa$, while the error (one standard deviation) on $\oo_{a}$ (and thus the detuning) is retrieved from a Gaussian fit to the sum of squared residuals, obtained by displacing the fit result with respect to the normal-incidence tranmission spectra of the antenna arrays.

\section{Finite-element simulations}
\label{sec:BS_FEM}
To gain more insight in our experimental results, and to calculate the different contributions to the complex frequency shift $\doo$, we perform 2D eigenfrequency simulations using the COMSOL software package (v5.0).
The cavity (diameter $=$ 36 $\um$) has a refractive index of 1.35, comparable to the experimentally determined effective index of our cavity mode and is placed 1.1 $\um$ from the substrate, which has a refractive index of 1.5.
The antenna is assigned a varying permittivity to mimic a change in detuning between cavity and antenna and for the surrounding we take a refractive index of 1.
The simulation includes the use of a Perfectly Matched Layer (PML).

\begin{figure}[tb]
\center
  \includegraphics[width=0.7\linewidth]{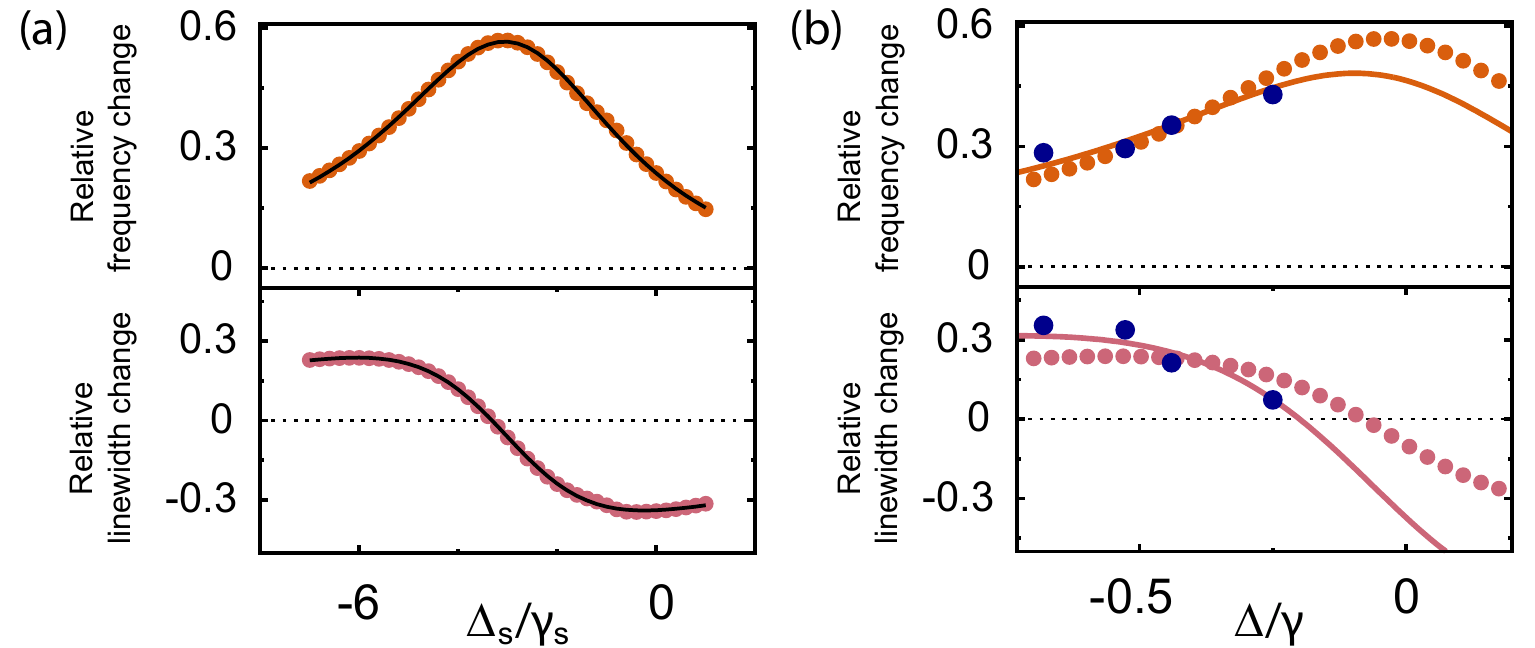}
  \caption{\textbf{Using multiple scatterers to find the scattering strength S.} a) Filled circles: change in resonance frequency (top) and linewidth (bottom) of the cavity when perturbed by 11 antennas. The values on the horizontal axis are the input parameters in the simulation, which do not include effects due to radiation losses. Performing a fit (black line) using a fano-lineshape allows us to obtain the corrected values $\Delta$ and $\gamma$, which are used in (b). b) The blue circles and solid lines are experimental data points and a fit using the coupled-mode model, respectively. The orange (top) and pink (bottom) filled circles are the relative frequency and linewidth change that are obtained from a simulation with S=340. The rescaled horizontal axis is obtained from the fit in (a).}
  \label{fig:comsol_11scatt}
\end{figure}

In our experiment, changing the length of the antennas results in a changing plasmon resonance, in turn causing a varying detuning between cavity and antenna resonance frequency. To mimic this in our simulation, we assign a tunable permitivitty, which has a Lorentzian lineshape, to the perturbing antenna(s):
\begin{equation}
\epsilon_{p} = 1- \frac{\text{S}*\gamma_{\mathrm{s}}/2}{\Delta_{\mathrm{s}}-\im\gamma_{\mathrm{s}}/2}
\end{equation}
Sweeping the variable $\Delta_{\mathrm{s}}$ now allows to calculate the change in cavity eigenfrequency as a function of detuning. The parameter S can be used to control the scattering strength of the antenna, while the antenna linewidth is held fixed at $\gamma_{\mathrm{s}}=1$. To determine a sensible value of S, we analyze the eigenfrequency shift of the cavity when perturbed by 11 antennas (pitch 1500,  Fig.~\ref{fig:comsol_11scatt}a) for different values of S and choose a value (S=340) which yields the best resemblance to our experiment (Fig.~\ref{fig:comsol_11scatt}b). The values for the relative change in linewidth and resonance frequency are obtained via similar analysis as performed in the experimental situation, thus involving separate simulations for the bare cavity, and cavity perturbed by just a glass substrate.

It is important to realize that the values $\Delta_{\mathrm{s}}$ and $\gamma_{\mathrm{s}}$ are input parameters in the simulation, which do not include the effect of radiation losses on the antenna linewidth and resonance frequency~\cite{SINovotny2012}.
To allow for a fair comparison with our experiment, we fit the results retrieved from the simulation with a Fano lineshape (black lines in Fig.~\ref{fig:comsol_11scatt}a), to obtain the radiation corrected values $\Delta$ and $\gamma$. Using these corrected values of $\Delta$ and $\gamma$, we rescale our horizontal axis and obtain Fig.~\ref{fig:comsol_11scatt}b.
Note that this approach resembles the experimental situation, where the radiation corrected values of $\Delta$ and $\gamma$ are retrieved from the normal-incidence transmission spectra.
It can be observed from Fig.~\ref{fig:comsol_11scatt}b that the orange (resonance frequency) and pink (linewidth) circles are a good, but not perfect, match to the experimentally obtained cavity frequency and linewidth change (blue circles).
We attribute the discrepancy to the intrinsic difference between a 2D simulation, where the cavity is modeled as an infinite cylinder, and the experimental situation, which employs a toroidal cavity.
This will necessarily change the radiation profile determining overlap and phase.
Nonetheless, the 2D simulation reproduces all the main features observed in the experiment.

\subsection{Cavity and perturbation without glass}
\begin{figure}[tb]
\center
  \includegraphics[width=0.7\linewidth]{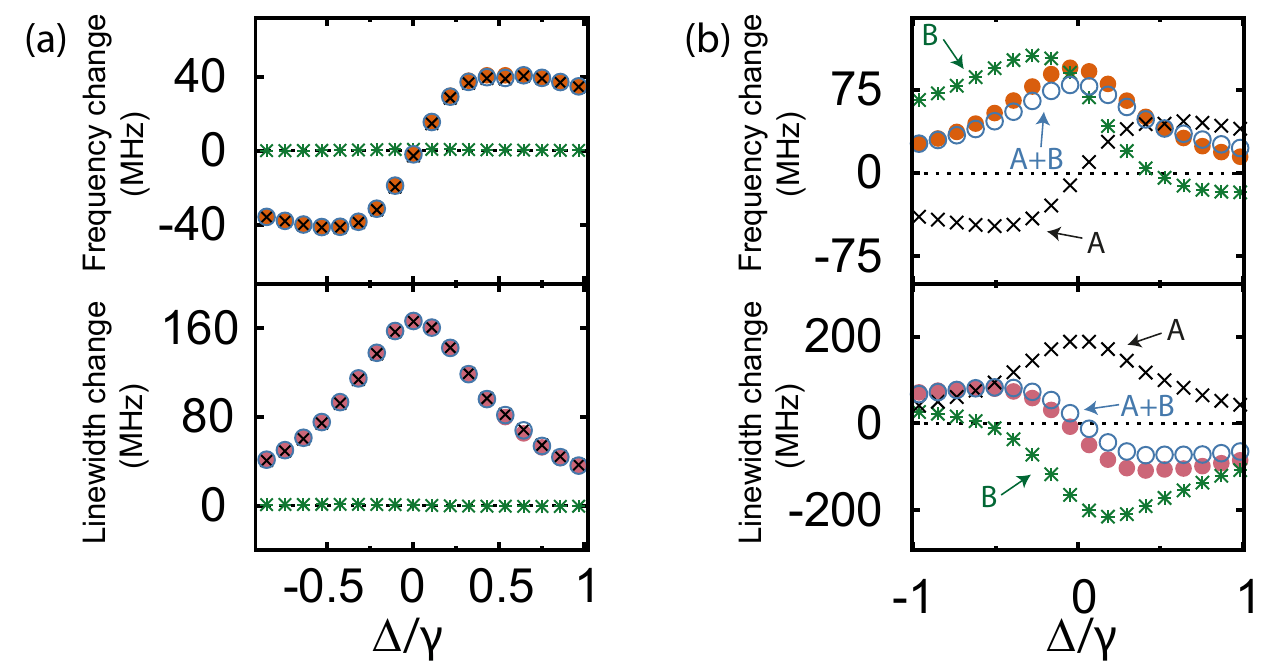}
  \caption{\textbf{Benchmarking the calculation.}
a) In the absence of a substrate, the cavity experiences a change in resonance frequency (top) and linewidth (bottom) that is solely associated with the polarizability (black crosses) of the perturbing particle, such that that the total eigenfrequency change as calculated by the Bethe-Schwinger equation (open blue circles) is completely dominated by local effects, leaving the radiative interaction (green stars) to be negligible. As expected,  the Bethe-Schwinger cavity perturbation formula is able to perfectly reproduce the cavity eigenfrequency shifts when compared to the shifts obtained using the eigenvalues (orange(pink) filled circles) calculated by COMSOL
b) In the situation where the substrate is present, as discussed in the main text, radiative interactions (green stars, A) are of importance when calculating the total (open blue circles, A+B) eigenfrequency shift of the cavity using the Bethe-Schwinger equation, dominating the local effect of the perturbation (black crosses, B). Note that the calculation of the full Bethe Schwinger equation yields almost perfect overlap with the cavity eigenfrequency changes as calculated by COMSOL (orange and pink filled circles), which are based on the eigenvalues of the solutions. We attribute the small difference that is observed mostly to the way joined data sets are handled in COMSOL, which involves projection on an intermediate mesh. This is known to cause loss of information and introduces a small error when integrating radiation patterns.}
  \label{fig:comsol_eigenvalues}
\end{figure}

When calculating the complete Bethe-Schwinger formula, it is useful to first consider the situation where a cavity is perturbed by a single perturbing particle, in the absence of a substrate, as this should not allow for radiation interaction taking place, and thus result in a perturbation of the cavity eigenfrequency which is solely governed by the polarizability  of the perturbing particle.
The results of this simulation are shown in Fig.~\ref{fig:comsol_eigenvalues}a.
It is clear that the cavity indeed lacks radiation interaction (green stars) with the antenna and experiences linewidth and resonance frequency shifts that are solely related to the polarizability of the particle (black crosses), such that the total eigenfrequency change (open blue circles) is completely dominated by local effects.
We remark that the Bethe-Schwinger cavity perturbation formula, which needs fields as it input, perfectly reproduces the cavity eigenfrequency shifts as predicted by the eigenvalues (orange(pink) filled circles) directly obtained from COMSOL.

\subsection{Cavity and perturbation with glass}
As discussed in the main text, introducing a glass substrate enables and extra coupling channel between antenna and cavity due to overlapping radiation profiles.
Fig.~\ref{fig:comsol_eigenvalues} shows results quantifying this effect for a cavity perturbed by a single antenna.
Here the radiative interactions (green stars) are of importance when calculating the total (open blue circles) eigenfrequency shift of the cavity using the Bethe-Schwinger equation, dominating the effect of local perturbation (black crosses).
Note that the calculation of complex $\doo$ using the full Bethe Schwinger equation, yields almost exactly the same results when compared to the cavity eigenfrequency changes based on the eigenvalues calculated in the simulation (orange(frequency) and pink(linewidth) filled circles).
We attribute the small difference that is observed mostly to the way joined data sets are handled in COMSOL.
The use of joined datasets is necessary to obtain the fields $\dE$, but causes an error when integrating the closed-surface flux integrals (last terms of equation~(\ref{eq:BS})).
The reason is that for joined datasets COMSOL projects full datasets on a less dense intermediate mesh~\footnote{COMSOL personal communication}.
We performed mesh convergence studies to see if this effect would diminish with increasing mesh density (up to 60 elements/$\lambda_{0}$).
Although we observe that both the eigenvalues and the fields of our solutions are clearly converged, the small difference in the cavity eigenfrequency shift that we obtain using both methods does not disappear.
Nonetheless, the discrepancy is small enough to reproduce all observed trends.

\subsection{Subtracting fields}
\begin{figure}[tb]
\center
  \includegraphics[width=0.7\linewidth]{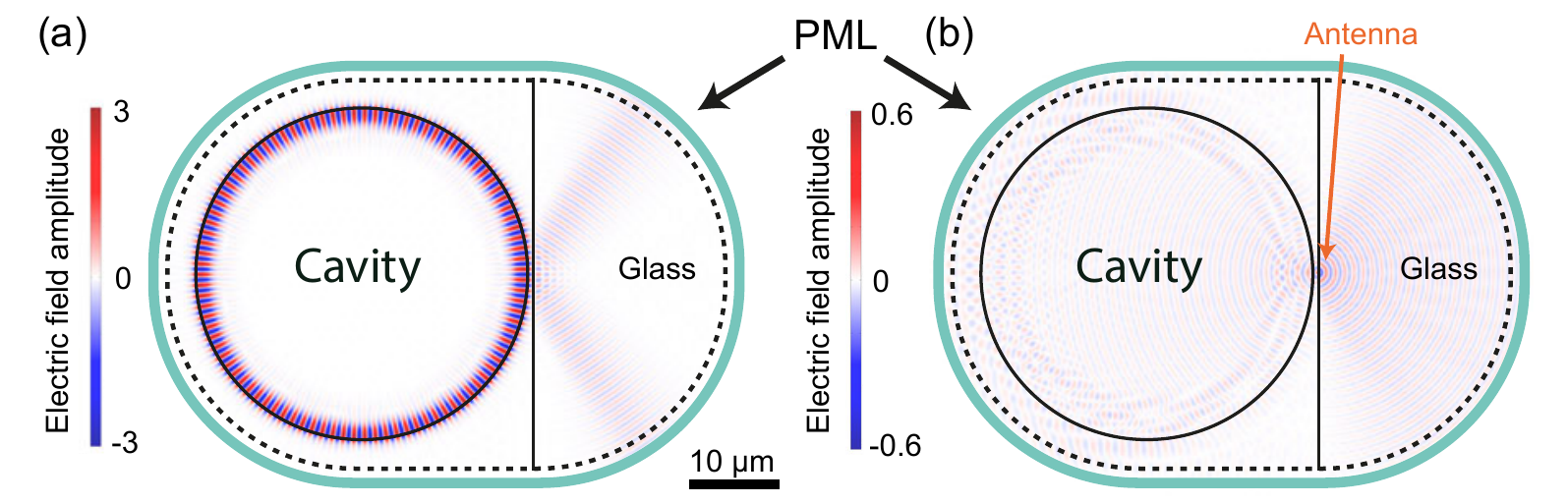}
  \caption{\textbf{Simulated field profiles.} a) The simulated field profile $\mathbf{E}_\text{0}$ for a cavity in the absence of perturbing particles. b) The field profile $\dE$ of the perturbing antenna. This profile is obtained by subtracting the fields in (a) from those shown in Fig.~\ref{fig:BS_comsol}a of the main paper. In both (a) and (b), the teal line surrounding the field profiles represents a perfectly matched layer (PML).   }
  \label{fig:comsol_field_profiles}
\end{figure}

Figure~\ref{fig:BS_comsol}a of the main paper shows the field profile $\mathbf{E}_\text{p}$ of a cavity perturbed by an antenna on a glass substrate.
To increase the visibility of the radiation pattern, that figure and Fig.~\ref{fig:comsol_field_profiles} show the field amplitude for a cavity placed 0.3 $\um$ from the substrate and antenna.
To calculate the full cavity perturbation formula, it is necessary to find the unperturbed cavity field profile (Fig.~\ref{fig:comsol_field_profiles}a, $\mathbf{E}_\text{0}$) and subtract this from $\mathbf{E}_\text{p}$ in order to obtain the field $\delta\mathbf{E}$ (Fig.~\ref{fig:comsol_field_profiles}b). The teal line surrounding the field profiles represents a perfectly matched layer (PML), which is used in all calculations discussed in this paper, while the dashed black line is the boundary which we used for the calculation of the surface integral of equation (\ref{eq:BS}).

A pitfall in subtracting fields from different eigenfrequency simulations lies in the arbitrary choice of amplitude and phase of the solutions. To overcome this problem, we monitor the amplitude and phase in both simulations (with and without antenna) in a point where modification of the field profile due to the perturbing antenna is expected to be negligible. This can be, for example, a point in the field maximum of the cavity mode far away from the perturbation, thus allowing to correct for possible differences in amplitude and phase between the calculations. To obtain the data in Fig~\ref{fig:BS_comsol}b, a single phase correction was performed, to match the phase at $\Delta_{\mathrm{s}}=-3$ to the unperturbed phase, in a mode maximum at large distance from the perturbing particle.

\section{Coupled mode theory}
\label{sec:coupled_oscillators}
We will describe the interactions between cavity and antenna using coupled mode theory~\cite{SISuh2004}, essentially modeling the system as coupled harmonic oscillators with Lorentzian response. We note that while it is possible to treat the antenna as a harmonic oscillator without making the Lorentzian approximation, all essential physics is captured in the simpler Lorentzian model. We treat the decay in a common (overlapping) radiation continuum, and its possible backaction on both antenna and cavity, by considering the `environment' as a third mode coupled to both antenna and cavity as well as to an independent decay port. The complex mode amplitudes of cavity ($a$), antenna ($b$) and environment ($c$) are normalized such that their magnitude squared equals the energy in each degree of freedom. They are collected in vector $\mathbf{a}=(a,b,c)^\mathrm{T}$. The equations of motion of the system (in absence of driving fields) can be written as
\begin{equation}
\frac{\mathrm{d}\mathbf{a}}{\mathrm{d}t} = \left(-\im \Omega - \Gamma\right)\mathbf{a},
\end{equation}
where
\begin{equation}
\Omega =
\begin{pmatrix}
\oc & -\gct^\ast & -\gkt \\
-\gct & \oo_0 & -\ggt^\ast \\
-\gkt^\ast & -\ggt & \oo_e
\end{pmatrix}
\end{equation}
is a Hermitian matrix containing the resonance frequencies and coupling rates. These coupling rates can in general be complex, and we will write them later as $\tilde{g}_j=g_j e^{\im\phi_j}$ with real $g_j$ and $\phi_j$. The matrix
\begin{equation}
\Gamma =
\begin{pmatrix}
\ko/2 & 0 & 0 \\
0 & \go/2 & 0 \\
0 & 0 & \zeta/2
\end{pmatrix}
\end{equation}
is also Hermitian and contains the damping rates to the different independent coupling channels, which are associated with an individual mode $a$, $b$, and $c$. As such, the output fields $\mathbf{s}=(s_a,s_b,s_c)^\mathrm{T}$ in these ports can be written as $\mathbf{s}=K\mathbf{a}$, where
\begin{equation}
K=
\begin{pmatrix}
\sqrt{\ko} & 0 & 0 \\
0 & \sqrt{\go} & 0 \\
0 & 0 & \sqrt{\zeta}
\end{pmatrix}.
\end{equation}
These equations satisfy time reversal symmetry and conservation of energy, assured by the hermiticity of $\Omega$ and $\Gamma$ together with the condition $K^\ast K = 2\Gamma$.
We now introduce the detunings $\Dc=\oo-\oc$, $\Delta=\oc-\oo_0$ and $\Dee=\oc-\oo_{e}$. The eigenmodes can be found by solving the system $M\mathbf{a}=\mathbf{0}$ where
\begin{equation}
M=I\oo-\Omega+\im\Gamma =
\begin{pmatrix}
\Dc+\im\ko/2 & \gct^\ast & \gkt \\
\gct & \Dc+\Delta+\im\go/2 & \ggt^\ast \\
\gkt^\ast & \ggt & \Dc+\Dee+\im\zeta/2
\end{pmatrix}.
\end{equation}
Since we assume the cavity and antenna are weakly coupled and we are only interested in the perturbed cavity mode, we can neglect $\Dc$ in the second and third row of $M$. The solution is
\begin{equation}
\Dc = \frac{\gk^2\left(\Delta+\im\frac{\go}{2}\right)+\gc^2\left(\Dee+\im\frac{\zeta}{2}\right)-2\gc\gk\gg\cos{\Phi}}{\left(\Delta+\im\frac{\go}{2}\right)\left(\Dee+\im\frac{\zeta}{2}\right) -\gg^{2}}-\im\ko,
\end{equation}
where we have introduced $\Phi=\phi_\mathrm{c}+\phi_\mathrm{\kappa}+\phi_\mathrm{\gamma}$.

It is instructive to consider the case where the cavity mode is absent ($a=0,\gc=\gk=0$). In that case, the power radiated via the environment into port $s_c$ is equal to $\zeta \gg^2 |b|^2/(\Dee^2+\zeta^2/4)$. We therefore define the rate
\begin{equation}
\gamma_1 = \frac{\zeta \gg^2}{\Dee^2+\zeta^2/4},
\end{equation}
which is the rate at which the antenna decays into the common radiation continuum if the cavity were not there. Likewise, we consider the case without the antenna ($b=0,\gc=\gg=0$), and define
\begin{equation}
\kappa_1 = \frac{\zeta \gk^2}{\Dee^2+\zeta^2/4},
\end{equation}
which is the rate at which the cavity would decay into this channel in absence of the antenna. In the latter case, the eigenmode would have detuning
\begin{equation}
\Delta_\mathrm{c}^{(b=0)} = \frac{\Dee}{\zeta}\kappa_{1} - \im\frac{\kappa}{2},
\end{equation}
where $\kappa=\ko+\kappa_1$ is the \emph{total} decay rate of the cavity without the antenna.
We now specifically write the general solution as $\Dc=\Delta_\mathrm{c}^{(b=0)}+\delta\Dc$, such that $\delta\Dc$ is the \emph{change} of the eigenmode frequency due to the presence of the antenna and $\delta\Dc=\doc-\im\delta\kk/2$.
Taking the limit of $\zeta$ to infinity, which effectively amounts to setting $\Dee=0$, it reads
\begin{equation}
\delta\Dc = \frac{\gc^2-\eta^2\kappa\gamma/4 + \im\gc\eta\sqrt{\kappa\gamma}\cos{\Phi}}{\Delta+\im\gamma/2},
\end{equation}
where we have introduced $\eta=\sqrt{(\kappa_1\gamma_1)/(\kappa\gamma)}$; a number between 0 and 1 that is related to the overlap of the radiation of both modes.
This is expression is equivalent to equation~(\ref{eq:3CO_Domega}) in the main text.

Next, let us turn to the relation of this model to the phase difference $\delta\phi$ as it appears in the Bethe-Schwinger equation. To this end, we express the field in the environment $c$, compared to the field in the cavity $a$. Its full expression (still taking $\Dee=0$) is
\begin{equation}
\frac{c}{a} = \frac{2\im e^{-\im\phi_\kappa}}{\zeta}\frac{\gk\left(\Delta+\im\frac{\gamma_0}{2}\right)-\gc\gg e^{\im\Phi}}{\Delta+\im\frac{\gamma}{2}}\label{covera}
\end{equation}
The phase $\delta\phi$ must be equal to the phase difference of the above in absence of the particle ($\gc=\gg=0$) and its phase in absence of direct coupling between cavity and environment ($\gk=0$). Performing this subtraction yields
\begin{equation}
\delta\phi = \pi+\Phi-\mathrm{arg}\left(\Delta+\im\frac{\gamma}{2}\right),
\label{eq:BSphi_toPhi}
\end{equation}
showing that the phase difference of the two decay paths is directly related to the (relative) phases of the coupling rates via $\Phi$ and the phase response of the antenna~\footnote{Due to the different (engineering) convention used in COMSOL, implementation of this formula requires to place a (-) sign in front of the right-hand side of expression~(\ref{eq:BSphi_toPhi}) when analyzing COMSOL data.}.
It becomes clear that the standard case, where the two paths interfere destructively on resonance ($\Delta=0$), occurs for $\Phi=\pi/2$, such that $\delta\phi=\pi$.

Finally, we note that it is straightforward to eliminate the environment mode $c$ from the model to construct an equivalent system of equations $M'(a,b)^\mathrm{T}=\mathbf{0}$. In that case the total damping rates $\kappa$ and $\gamma$ of the uncoupled cavity and antenna appear on the diagonal:
\begin{equation}
M'=
\begin{pmatrix}
\Dc+\im\frac{\kappa}{2}& e^{-\im\phi_\mathrm{c}}\left(\gc+\frac{\im}{2}\sqrt{\kappa_1\gamma_1}e^{\im\Phi}\right)  \\
e^{\im\phi_\mathrm{c}}\left(\gc+\frac{\im}{2}\sqrt{\kappa_1\gamma_1}e^{-\im\Phi}\right) & \Delta+\im\frac{\gamma}{2}
\end{pmatrix}.
\end{equation}
We can choose the phase $\phi_\mathrm{c}=0$ without loss of generality, such that
\begin{equation}
M'=
\begin{pmatrix}
\Dc+\im\frac{\kappa}{2}& \gc+\frac{\im}{2}\sqrt{\kappa_1\gamma_1}e^{\im\Phi}  \\
\gc+\frac{\im}{2}\sqrt{\kappa_1\gamma_1}e^{-\im\Phi} & \Delta+\im\frac{\gamma}{2}
\end{pmatrix}.
\end{equation}
Note that the off-diagonal matrix elements of $M'$, representing coupling between cavity and antenna, are now no longer Hermitian.

\section{Varying antenna pitch}
\label{sec:BS_pitches}
\begin{figure}[tb]
\center
  \includegraphics[width=0.4\linewidth]{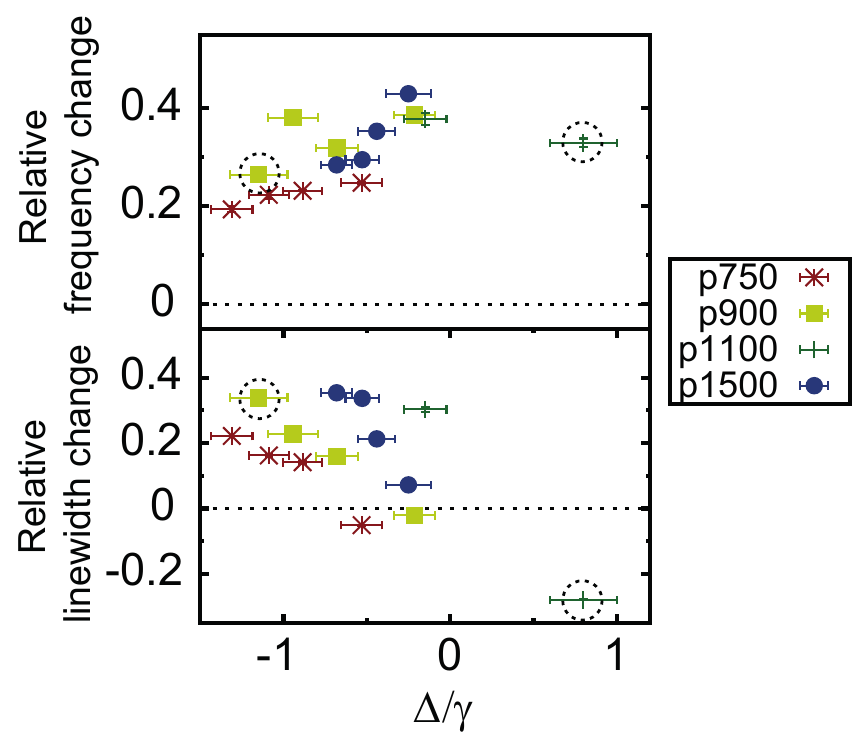}
  \caption{\textbf{Varying the antenna array pitch.} Changing the pitch along the antenna short axis (between 750 and 1500 nm) does not greatly affect the cavity eigenfrequency shift. For all arrays, we observe cavity blueshifts (top panel) and linewidth narrowing (bottom panel). Note that in the bottom panel, the different pitches cross the point of zero linewidth change at slightly different values of $\Delta/\gamma$. Modification of the single particle polarizability due to the array is the most likely cause of this behaviour. The values on the vertical axis are corrected for the unit-cell area and the points which are surrounded by a dashed circle correspond to the arrays used to generate Fig.~\ref{fig:BS_exp}a of the main text. Errors on the mean (vertical axis) fall mostly within the plot markers and the error on the horizontal axis is one standard deviation, retrieved from a Gaussian fit to the sum of squared residuals (obtained by displacing the fit result). Due to limited detector range, a reliable horizontal error for the data at P1100,$\Delta/\gamma\approx 0.8$ could not be obtained using this method. The error on this data point was thus set equal to the error for array P1100,$\Delta/\gamma\approx-0.15$.}
  \label{fig:varying_pitch}
\end{figure}

The text of the main paper shows the results and analysis of antenna arrays with pitch 800(1500) nm along the long(short) axis of the antennas.
More pitches are, however, experimentally investigated and Fig.~\ref{fig:varying_pitch} shows the main results which we obtain when varying the pitch along the short axis (750-1500 nm) of the antennas, while the pitch along the antenna long axis is kept constant at 800 nm.
In this figure, the points which are surrounded by a dashed circle correspond to the arrays used to generate Fig.~\ref{fig:BS_exp}a of the main text.
The main conclusion that can be drawn from Fig.~\ref{fig:varying_pitch} is that, qualitatively, the different arrays behave the same: all arrays induce blueshifts of the cavity resonance frequency (top panel) and linewidth narrowing (bottom panel) of the cavity on approaching $\Delta/\gamma=0$.
With changing pitch, the value of $\Delta/\gamma$ where the "relative linewidth change" equals 0, slightly varies.
We attribute this effect to the influence of the array on the single particle polarizability.
Remarkably, for the largest possible detuning we measure linewidth narrowing up to 30\% for a pitch of 1100 nm.

As in the main paper, the values of $\Delta/\gamma$ are retrieved from a fit of the normal-incidence transmission spectra using a Lorentzian lineshape.
Lattice sum calculations~\cite{SIDeAbajo2007} are performed to exclude strong modification of $\Delta$ due to the change of effective angle of incidence (in the experiment) with respect to the normal-incidence transmission spectra.
In addition, total internal reflection measurements on the samples with a 1500 nm pitch showed no significant change of $\Delta$ with respect to the values obtained from the normal-incidence measurements.

\end{document}